\RequirePackage{fix-cm}
\documentclass[a4paper,man,natbib,floatsintext]{apa6}
\usepackage[english]{babel}
\usepackage{graphicx,subcaption}
\usepackage{etoolbox} 
\usepackage{tikz}
\usetikzlibrary{shadings,positioning,fit,calc,shapes.geometric}
\usepackage{standalone}
\usepackage{amsmath}
\PassOptionsToPackage{hyphens}{url}\usepackage[hidelinks=true]{hyperref}
\hypersetup{
  colorlinks   = true, 
  urlcolor     = black, 
  linkcolor    = black, 
  citecolor   = black 
}

\usepackage[colorinlistoftodos,prependcaption,textsize=medium]{todonotes}
\usepackage{xargs}
\usepackage{xstring}
\newcommandx{\ny}[2][1=]{\todo[linecolor=red,backgroundcolor=red!25,bordercolor=red,#1,inline]{NY: #2}} 
\newcommandx{\vk}[2][1=]{\todo[linecolor=blue,backgroundcolor=blue!25,bordercolor=blue,#1,inline]{VK: #2}} 
\newcommandx{\kt}[2][1=]{\todo[linecolor=green,backgroundcolor=green!25,bordercolor=green,#1,inline]{KT: #2}} 
\newcommandx{\ac}[2][1=]{\todo[linecolor=brown,backgroundcolor=brown!25,bordercolor=brown,#1,inline]{AC: #2}} 

\usepackage{mathptmx}      

\newcommand{\framework}{Delayed Feedback based Immersive Navigation Environment }
\newcommand{\ack}{DeFINE} 

\graphicspath{{./sections/figures/}} 
\newrobustcmd*{\mytriangle}[1]{\tikz{\filldraw[draw=#1,fill=#1] (0,0) --
(0.2cm,0) -- (0.1cm,0.2cm);}}

\colorlet{mygreen}{green!80!black}
\colorlet{myblue}{blue!80!black}
\colorlet{myred}{red!80!black}

\usepackage{ntheorem}

\makeatletter
\newcounter{subhyp} 
\let\savedc@hyp\c@hyp

\newcommand{\normhyp}{%
  \let\c@hyp\savedc@hyp 
  \renewcommand\thehyp{\arabic{hyp}}%
} 
\makeatother

\abstract{With the advent of consumer-grade products for presenting an immersive virtual environment (VE), there is a growing interest in utilizing VEs for testing human navigation behavior.
However, preparing a VE still requires a high level of technical expertise in computer graphics and virtual reality, posing a significant hurdle to embracing the emerging technology.
To address this issue, this paper presents \framework (\ack), a framework that allows for easy creation and administration of navigation tasks within customizable VEs via intuitive graphical user interfaces and simple settings files.
Importantly, \ack~has a built-in capability to provide performance feedback to participants during an experiment, a feature that is critically missing in other similar frameworks.
To show the usability of \ack~from both experimentalists' and participants' perspectives, a demonstration was made in which participants navigated to a hidden goal location with feedback that differentially weighted speed and accuracy of their responses.
In addition, the participants evaluated \ack~in terms of its ease of use, required workload, and proneness to induce cybersickness.
The demonstration exemplified typical experimental manipulations \ack~accommodates and what types of data it can collect for characterizing participants' task performance.
With its out-of-the-box functionality and potential customizability due to open-source licensing, \ack~makes VEs more accessible to many researchers.


}

\begin{document}

\title{DeFINE: \framework for Studying Goal-Directed Human Navigation}
\shorttitle{DeFINE}

\threeauthors{Kshitij Tiwari* and Ville Kyrki}{Allen Cheung}{Naohide Yamamoto*}
\threeaffiliations{Department of Electrical Engineering and Automation, Aalto University}{Queensland Brain Institute, The University of Queensland}{School of Psychology and Counselling, Queensland University of Technology (QUT)}
 
\authornote{Orcid IDs: Kshitij Tiwari \url{https://orcid.org/0000-0003-1789-7961}
	
	\hspace{55pt}Ville Kyrki \url{https://orcid.org/0000-0002-5230-5549}
	
	\hspace{55pt}Allen Cheung \url{https://orcid.org/0000-0001-9770-217X}
	
	\hspace{55pt}Naohide Yamamoto \url{https://orcid.org/0000-0001-9734-7470}
	
	The authors would like to thank Onur Sari and Ville Sinkkonen for their contributions to the development of the framework presented in this article.

	*Corresponding authors---KT: Center of Ubiquitous Computing, Faculty of Information Technology and Electrical Engineering, University of Oulu, Finland (email: kshitij.tiwari@oulu.fi); NY: School of Psychology and Counselling, Queensland University of Technology, Brisbane, Australia (email: naohide.yamamoto@qut.edu.au).
	}

\maketitle

Behavioral researchers are increasingly becoming interested in understanding the underlying mechanisms for goal-directed navigation in humans~\citep{spiers2006thoughts,cornwell2008human,pezzulo2014internally}. 
Whilst it would be ideal to gather data on navigation in a well-controlled real-world setting, not every researcher has access to such a setup. 
However, with the advances in virtual reality (VR), it is becoming easier to set up economical game-like environments to test hypotheses about human navigational behaviors.


Although VR technologies are becoming more accessible, they are primarily for playing computer games and not for carrying out behavioral experiments.
Thus, to use them for investigating human navigation, researchers need to have working knowledge of graphics design and game engines like Unity (\url{https://unity.com/}) and Panda3D (\url{https://www.panda3d.org/}), which may not be the case even when they are proficient in general scientific programming. 
To address this issue, attempts have been made at developing easy-to-use VR frameworks for researchers who are novice at computer graphics and game creation. 
Notable examples include Python-based Experiment Programming Library \citep[PyEPL; ][]{geller2007}, Maze Suite \citep{ayaz2008,ayaz2011}, PandaEPL \citep{solway2013}, Experiments in Virtual Environments \citep[EVE; ][]{grubel2017}, Virtual Reality Experiments \citep[VREX; ][]{vasser2017vrex}, VRmaze \citep{machado2019new}, Unity Experiment Framework \citep[UXF;][]{Brookes2019}, Route Learning and Navigation Test Battery \citep{wiener2019}, NavWell \citep{commins_2019}, BiomotionLab Toolkit for Unity Experiments \citep[bmlTUX;][]{bebko2020}, and Landmarks \citep{starrett2020}.

These existing frameworks are similar in that they offer some or all of the following functions for conducting behavioral experiments in VR: modeling and rendering virtual environments (VEs), designing the structure of an experiment (e.g., setting the number and order of trials), executing the trials, recording data (e.g., participants' navigation trajectories), and performing preset analyses of the recorded data.
A major difference between the frameworks is in the extent to which they are designed for specific purposes.
For example, Maze Suite is specialized for creating and running experiments in standard mazes (i.e., mazes created by dividing an enclosed space into connected paths by walls).
By limiting its scope, Maze Suite achieves a high level of ease of use---that is, everything can be done in graphical user interfaces (GUIs) and no coding is necessary on the part of end-users.
At the other end of the spectrum, PyEPL and PandaEPL are Python libraries for programming behavioral experiments in general (PyEPL) and spatial navigation experiments in particular (PandaEPL).
Because they are not compiled software packages but programming libraries, researchers can create any experiments using PyEPL and PandaEPL, as far as their programming proficiency goes.
The other frameworks lie in between, balancing ease-of-use and flexibility to varying degrees by preparing some ready-made functionality while providing users with the way of doing their own coding for fine customization.
Some of them are still relatively specific to certain types of experiments (e.g., VREX focuses on indoor environments that consist of connected rooms).
The others offer generic modules for experimental design and data recording, leaving high-level features of experiments including stimulus presentation up to users' programming (e.g., UXF). 

Despite the different purposes these frameworks are designed to achieve, one aspect shared by them is that
%
%
they focus on the stimulus--response relationship in examining navigational behavior. 
That is, participants are presented with a stimulus with which they carry out a navigation response (e.g., walking to a goal location indicated by visual cues), and during and at the completion of the response, they do not receive any feedback on their performance.
Although such a research design is appropriate and even required for investigating certain aspects of navigational behavior~\citep{philbeck_comparison_1997,loomis_visual_2003}, it makes it impossible to examine how participants modulate their subsequent response to the stimulus using feedback~\citep{brand2008does}. 

Importantly, when goal-directed navigation is performed in real-world settings, navigators often receive feedback with which they can adjust their behavior.
For example, when they walk in the dark to reach a door and fail to touch its knob, the lack of tactile sensation serves as feedback and informs them that they still have a few more steps to go. 
In this instance, the navigational behavior should be characterized as consisting of a closed stimulus--response--feedback loop, instead of an open stimulus--response loop as in the foregoing frameworks. 
Indeed, studies have been conducted to capture human navigation by the stimulus--response--feedback loop.
For instance, \cite{carton2016towards} have found evidence that humans tend to adapt their trajectory planning horizon (i.e., how far in the future they plan their locomotion trajectory) when they detect potential collision situations. 
In this case, impending collisions constitute a stimulus that triggers a response, which is witnessed in the form of a shortened planning horizon. 
The successful avoidance of collision or a failure thereof functions as feedback that can subsequently be used to tune the length of planning horizons.
In this manner, experiments on human navigation come in a variety of designs, utilizing both stimulus--response and stimulus--response--feedback loops.
To fully accommodate this diversity of the experiments, there is a need for a new VR framework that allows for incorporation of feedback into the experimental design.
To this end, we developed the \framework~(\ack).
\section{\ack: \framework}

\ack~is freely available to download via GitHub at \url{https://github.com/ktiwari9/define-VR}.
\ack~is based on the Unity game engine, and hence, relies heavily on C\# as a programming language. 
All the low-level implementation is already taken care of to minimize the workload of end-users who will use \ack~in its default settings or with minimal customization as required by their experimental design.
\ack~aims specifically to provide an easy-to-use experimental framework that is based on the stimulus--response--feedback architecture, which can be used to study goal-directed spatial navigation in an immersive three-dimensional space.
In order to reduce the burden of researchers when setting up an experiment, 
\ack~allows them to make simple alterations of experimental parameters through GUIs and by changing settings files that are in a straightforward JSON format.
We provide a short video that succinctly explains how to use the basic functionality of \ack~``out-of-the-box'' (\url{https://youtu.be/OVYiSHygye0}).
To allow for further customization of the experiment and the framework itself, \ack~is being released open-source under the MIT license.
Thus, interested researchers can modify its code directly.
To facilitate this, \ack's codebase is made modular, making it possible to alter a particular pre-programmed functionality by adjusting the code in a specific module only, and also to incorporate a new functionality into the framework by creating a new module and placing it in an appropriate folder. 
%
The file hierarchy is kept intuitive for this purpose.
Another video we provide offers a quick tutorial of how to change various elements of \ack~using the Unity software (\url{https://youtu.be/smIp5n9kyAM}).
A detailed user manual is also available (\url{https://github.com/ktiwari9/define-VR/blob/master/user_manual.pdf}).

Currently, \ack~can be integrated into Unity tasks built for Windows personal computers. 
It is assumed that \ack~will be used with a head-mounted display (HMD) such as HTC Vive and Oculus Rift.
For example, \ack~is designed to utilize the HMD's motion tracking sensors for implementing various methods of participants' locomotion within VEs (see the locomotion methods section below for details).
In addition to the HMD worn by the participants, \ack~simultaneously presents a VE to a desktop display so that experimentalists can monitor the progress of an experiment.
Further details about hardware and software requirements for \ack~are available in the user manual.

The main capabilities and options of \ack~are detailed below in the following order: (1) the generic experimental structure, (2) time- and accuracy-based feedback, (3) the GUI, (4) a diverse suite of locomotion methods, (5) static and dynamic goals, (6) performance leader-board, and (7) intra-VR surveys.

\subsection{Experiment Structure}

\begin{figure}[!hb]
	\centering
	\includegraphics[scale=0.5]{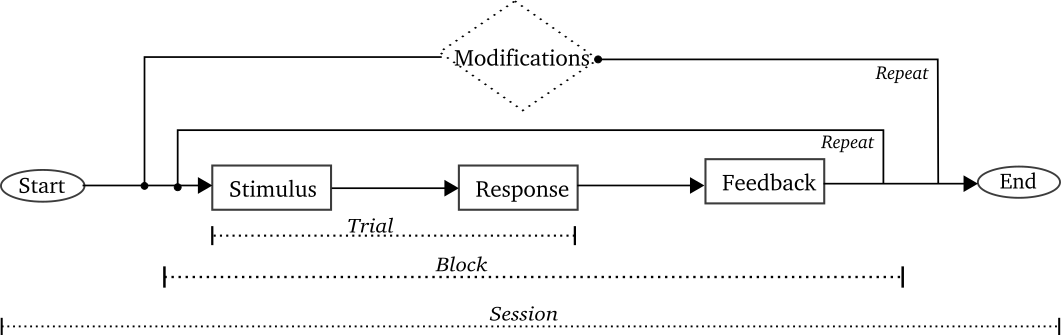}
	\caption{Structure of typical human spatial navigation experiments. 
	Participants are presented with a \textit{stimulus}, to which they provide a \textit{response} and are given \textit{feedback} after the \textit{trial} has been fully executed. 
	This \textit{feedback} is then used to modify performance in the next trial.
	Usually, multiple trials are carried out under the same condition, which constitute a \textit{block}. 
	Once the participants are finished with the block, they progress to the next block which may involve modifications of the condition, and their navigation performance is evaluated again over multiple trials. 
	A single iteration of the task is called a \textit{session}.
	This trial--block--session architecture as well as the diagram itself are modeled after those presented by \cite{Brookes2019}.}
	\label{fig:loop}
\end{figure}

Human behavioral experiments are often defined by a \textit{trial--block--session} architecture which allows experimentalists to repeat a task multiple times to acquire requisite data (Figure~\ref{fig:loop}). 
Just like the UXF~\citep{Brookes2019}, \ack~adopts this architecture.
A \textit{trial} is an instance of the experiment where participants are presented with a stimulus and their response is recorded. 
At the end of the trial, the participants receive feedback.
To clarify that the feedback is given after the response is made, as opposed to during the trial as the response unfolds, this feedback is referred to as ``delayed'' in \ack. 
Trials are often repeated multiple times for various purposes (e.g., to measure variability of responses to the same stimulus, to decipher the learning effects over trials, or to train the participants on the task), constituting a \textit{block}.
At the end of the block, the participants are assumed to be familiar with the environment and to have formulated a behavior of choice for the task at hand. 
In order to evaluate the quality of this behavior, the experimentalists may choose to make some modifications to the environment before proceeding with the next block. 
The experiment can consist of a single or multiple blocks, and when there are multiple blocks, a single iteration of the task over the blocks is called a \textit{session}.

\begin{figure}[!htbp]
	\centering
	\includegraphics[scale=0.7]{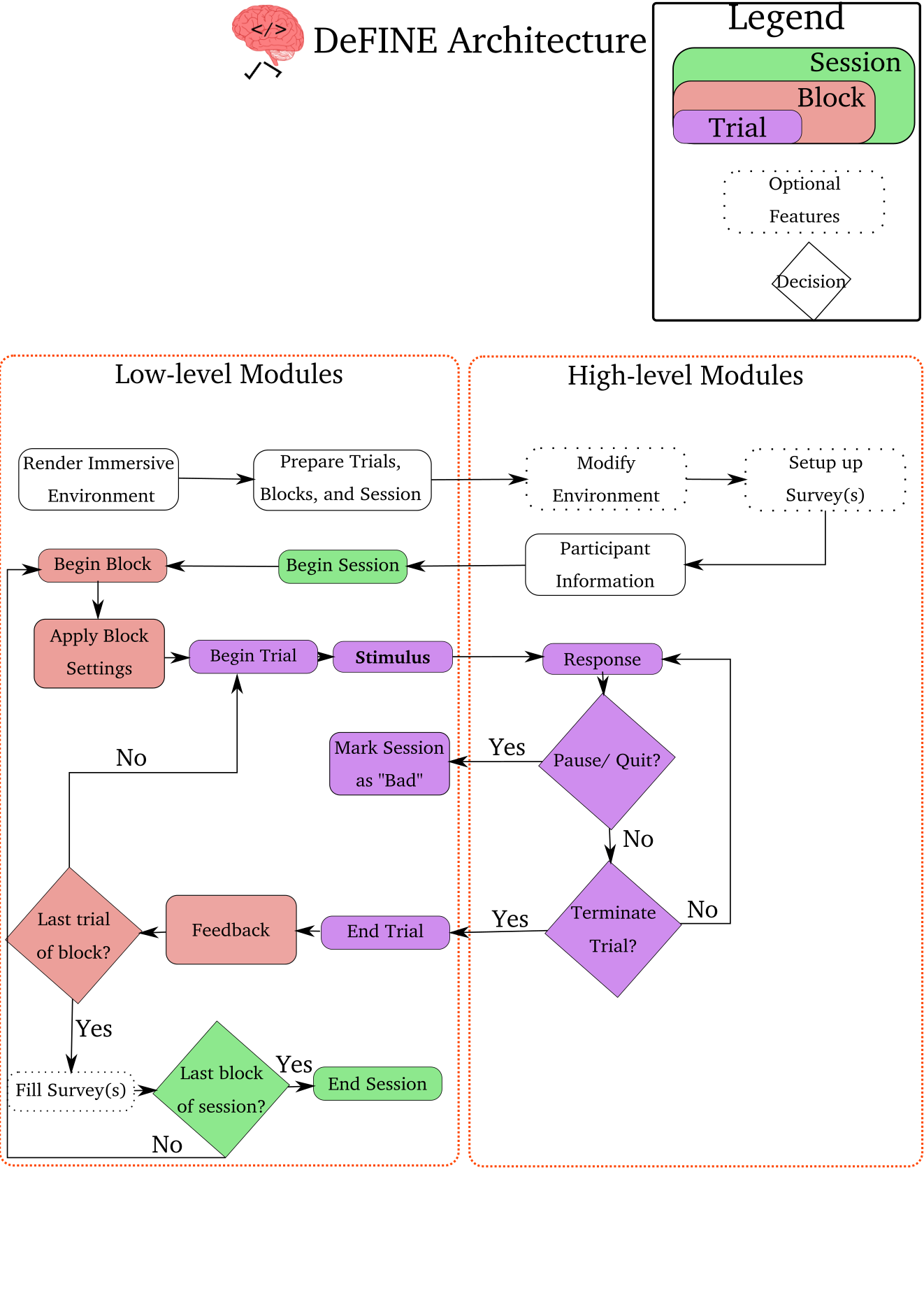}
	\caption{The architecture of the \framework~(\ack). 
	The left panel shows the low-level functionality provided by \ack~while the high-level experimentalist-defined functionalities are shown on the right. 
	The surveys/questionnaires are optional, and hence, shown using dashed lines. 
	Modules shown in purple are parsed during a \textit{trial}, those in red during a \textit{block} and those in green during a \textit{session}.
	The logo of \ack~shown at the top was inspired by that of UXF \citep{Brookes2019}, symbolizing that \ack~is built as a navigation-focused extension of UXF.}
	\label{fig:arch}
\end{figure}

\ack~has primarily two levels of abstraction as shown by the two columns (demarcated by orange dotted lines) in Figure~\ref{fig:arch}.
All modules mentioned on the left are those implemented at the low-level abstraction that come pre-programmed with \ack, while those on the right are implemented at a higher level for easy and quick modifications.
Such an arrangement streamlines customization of an experiment to suit different research studies because the modules at the low level of abstraction are common across many experiments and thus, the experimentalists can rely on \ack's built-in functionality, which in turn allows them to focus their effort on customizing high-level modules that tend to be unique to each experiment. 
All modules shown by dotted black lines represent optional modules that can be utilized as seen fit by the experimentalists. 
In keeping with the trial--block--session architecture as described in Figure~\ref{fig:loop}, all modules clustered under the green polygon (``Begin Session'') represent a session, those under the red polygon (``Begin Block'') represent blocks, and those under the purple polygon (``Begin Trial'') represent trials. All other modules (under either column of abstraction) are preset before the session starts. 

The flow of an experiment starts from the top of Figure~\ref{fig:arch}. 
\ack~takes care of initializing and rendering an immersive environment and setting up a session, blocks, and trials, but the experimentalists are required to configure the VE (e.g., setting its size) and specify the experimental design (e.g., the number of trials per block) by modifying the settings files in order to make \ack~behave in the way they need for their experiment. 
Details of the settings files are described later.
If the experimentalists want questionnaires to be filled within the VE, these questionnaires need to be created online, and the respective links to the questionnaires need to be added to the environment settings. 
Setting up the questionnaires is followed by inputting a participant's information and selecting preset settings that are appropriate for the experiment. 
Once the participant information has been entered, the experimentalists can start the experimental session that \ack~will generate. 
If there are block-specific settings, those are applied by \ack~before starting the first trial. 
The trials follow the stimulus--response--feedback structure, by showing the participant a stimulus in the environment, to which the participant responds, and after the experimentalist-defined end condition has been met, the trial ends and feedback is shown to the participant. 
A new trial is started automatically after the feedback has been given, until the program reaches the end of the block. 
At the end of the block, the participant will be shown the questionnaires, if any were specified during the initial setup. 
During the trials the experimentalists can take down notes, or manually mark the session as bad to indicate that it should not be taken into account for data analysis. 
At any point of the session the experimentalists can abort the session, which will also mark the session as bad in the stored session notes. 
If there are more blocks remaining in the session, the next block is started after \ack~has applied its specific settings to the environment. 
After the final block of trials has been completed, and the questionnaires following it have been answered, the session ends and the participant and experimentalists are returned to the startup screen. 
For each of the trials executed with \ack, the participant's trajectory during the trial, the status and changes of environment variables, the feedback score obtained at the end of the trial (see the next section for details), trial start time, trial end time, total time taken for the trial, and straight-line distance to the goal when the participant ends the response are logged.  

\subsection{Feedback}
As opposed to other closed-loop systems where real-time feedback may be made available to participants as they carry out a response, \ack~provides delayed feedback at the end of each trial.
This is because continuous feedback will most likely go against the purpose of typical goal-directed navigation experiments.
Usually, these experiments are to test participants' ability to estimate their location relative to a goal by using sensory and other spatial cues in an environment \citep{loomis_measuring_2008}.
If the participants were given external feedback about whether they were moving in the right direction for every step they took, this feedback would essentially be a non-spatial cue that would directly aid them in their location estimation.
In extreme cases, with continuous feedback, the participants could perform the task by moving in an arbitrary direction without processing the sensory and spatial cues and seeing if that would result in positive feedback.
Such a strategy would lead them to take myopic unstructured paths to the goal, causing non-optimal navigation performance. 
Thus, by default, \ack~is designed to give performance feedback only after the trial is completed.
However, it is possible for experimentalists to modify \ack's source code in Unity and have it provide real-time feedback, if they so choose.

Feedback on a goal-directed navigation behavior can be given in a number of different forms, but \ack~adopts a reward/cost function that evaluates participants' performance and provides feedback as gains and losses of scores.
It has been shown that feedback of this type is very effective in affecting participants' behavior and decision making under a variety of conditions \citep{brand2008does,hossain_behavioralist_2012,yechiam_loss_2014}. 
The reward and cost in the context of navigation can also be defined in various ways, and it is up to experimentalists' discretion how they formulate the reward/cost function in \ack, but one straightforward method would be to define them by using speed and accuracy of navigation.
That is, the quicker the participants are in performing a trial, the greater the reward (or the smaller the cost); and the more accurate they are in reaching a goal, the greater the reward (or the smaller the cost).
By default, \ack~implements a reward/cost function of this form. Specifically:
\begin{equation}
R = \beta_1 \exp (-\alpha_1 t) + \beta_2 \exp (-\alpha_2 d)
\label{eqn:reward_func}
\end{equation}

In Equation~\eqref{eqn:reward_func}, $t$ refers to the time taken for navigating towards a goal in a trial, and $d$ refers to the residual straight-line distance to the goal from the location at which participants end the trial. 
$\beta_1\text{ and }\beta_2$ are weights used for combining the time and distance into the decaying reward function, which penalizes both the time taken and the residual distance to the goal (i.e., rewarding shorter times and smaller distances with higher scores).
$\alpha_1\text{ and }\alpha_2$ are factors for scaling the effects of time and distance. 
If experimentalists choose to use this function in their experiments, they can assign values of their choice to these parameters simply by specifying them in a settings file (details shown later).
If they are to calculate the reward/cost scores using their own equation, they can do so by modifying a relevant section of \ack's codes in Unity.
It should further be noted that, by changing the relevant codes and implementing their own equation, the experimentalists can use any kinds of feedback that do not take the form of a cost/reward function.
For example, it is possible to simply present how far away participants are from the goal at the end of each trial.

\subsection{Graphical User Interface (GUI)}
\begin{figure}[!htbp]
	\begin{subfigure}{.5\textwidth}
	\centering
	\includegraphics[scale=0.6]{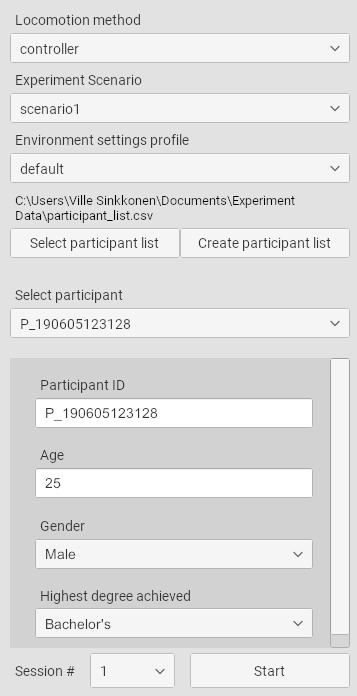}
	\caption{Drop-down menus to specify experimental parameters \citep[adapted from UXF;][]{Brookes2019}.}
	\label{fig:GUI}
\end{subfigure}%
\begin{subfigure}{.5\textwidth}
	\centering
	\vspace{9cm}\includegraphics[width=0.5\textwidth]{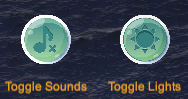}
	\caption{The buttons for toggling sound and lights.}
	\label{buttons}
\end{subfigure}
\caption{\ack's intuitive GUI for experimentalists to set up an experiment.}
\end{figure}

Utilizing the GUI of UXF~\citep{Brookes2019}, \ack~allows experimentalists to log participant information including, but not limited to, name (or participant identification), age, gender, and educational qualification (Figure~\ref{fig:GUI}). 
Should other personal particulars be required, they can be easily added to the framework by modifying relevant settings files. 
As an extension to UXF's original GUI, \ack~ allows the experimentalists to quickly set up the environment of choice with a desired locomotion method (see the next section for details about the locomotion methods). 
This unique feature can also be scaled and automated to handle multiple combinations of environments and locomotion methods via \ack's auto-pilot mode. 
In this mode, the experimentalists can provide \ack~with preset instructions so that it loads specific combinations of the environment and the locomotion method in a specified order.
This way, a sequence of participants can be tested automatically, doing away with the need to individually set up an appropriate combination of the environment and the locomotion method for every participant. 
For example, if an experimental design requires that each participant is shown a different environment, a sequence of environments can be explicitly listed in settings files which will then be autonomously parsed when executing the auto-pilot mode. 
If each participant is to do trials with a different locomotion method, explicit participant-locomotion method combinations can be listed in the settings files in a similar fashion.

As a significant extension to the predecessor, UXF, \ack~also provides functionalities to study the role of lighting conditions and auditory cues in spatial navigation. 
At any point during a trial, experimentalists can toggle the lights and sound of a VE on or off by clicking on the dedicated buttons on the user interface, shown in Figure~\ref{buttons}. 
The change of the status of these environmental variables are logged along with the information about participants' performance in a navigation task (e.g., their position within the environment at a given time point).

\subsection{Locomotion Methods}
In order to provide a locomotion suite for participants to perform goal-directed navigation in VR, \ack~comes equipped with a variety of locomotion methods.

\subsubsection{Teleoperated locomotion} 
In order to allow teleoperated locomotion, \ack~is compatible with both a keyboard based and a VR controller based teleoperated methods. 
A typical use case may involve head direction sensors in an HMD being used to update participants' headings in a VE while the keyboard or the VR controller being used to linearly traverse at preset velocity (which is to be specified in a settings file by experimentalists).
The necessary key-bindings and further details are available in the user manual (\url{https://github.com/ktiwari9/define-VR/blob/master/user_manual.pdf}).

\subsubsection{Arm-swing locomotion}
Arm-swing locomotion is an implementation of walking-in-place locomotion. 
In this method, the participants walk in place, including swinging their arms in a manner consistent with their pace of walking. 
It has been shown that such arm swings are effective in having the participants experience a naturalistic sense of locomotion without actually moving in real space \citep{kunz_evidence_2009,yamamoto_imagined2018}.
This locomotion method uses the physical movement of the VR controller(s), held by the participants, to determine forward speed in the VE. 
The movement speed is calculated from the positional difference of the tracked controller(s) between two consecutive frames. 
This calculation can be done either by requiring movement of two controllers (typically, one in each hand) or by using one controller (or either of the two controllers) that moves more than a given threshold amount between the frames. 
When the use of the two controllers is required, the forward speed in the environment is set to be zero, unless both of the controllers exceed the threshold value.

\subsubsection{Head-bob locomotion}
Head-bob locomotion is another implementation of walking-in-place locomotion. 
In order to move forward in the VE, the participants need to walk in place, and as they do, their head, and in particular, the HMD, bobs slightly vertically. 
This locomotion method uses this vertical bobbing to determine the forward velocity.
\ack~tracks the vertical direction of the bobbing and its starting position. 
Once the direction changes, the participants are considered to have stepped, if the vertical height difference between two successive flexion points exceeds a threshold value specified in the settings of the locomotion method. 
The detected physical step is then translated into a step in the VE so that the participants walk forward in the VE at preset velocity.
Due to the fact that the HMD is in front of the participants' face, turning their head up or down causes the HMD to move vertically. 
In order to avoid reading these vertical movements as steps of the participants, \ack~also tracks the participants' rotational head movements about the pitch axis and ignores any ``bobs'' that are accompanied with the rotational head movements that exceed a specified threshold value.

\subsubsection{Physical walking}
Physical walking is the only locomotion method in which the participants physically move around in the real world. 
The movement of the participants is tracked by using the HMD's motion tracking sensors and the participants' position in the VE is updated accordingly.
Owing to the limited size of a physical area in which the participants' movement can be tracked (which is typically around $10 \times 10$~m), the size of the VE is going to be limited.
To alleviate this limitation, sometimes modified physical walking methods such as redirected walking are adapted~\citep{paludan2016disguising,nilsson201815}.
In these methods, the rotations and translations of the participants are slightly altered between physical and virtual worlds in order to steer the participants away from the edges of the available physical area.
However, \ack~does not utilize these methods because they can induce disruption to mental and neural spatial representations as well as to navigational behavior by causing a mismatch between intended (and physically carried out) movements and consequent virtual movements \citep{du_unidirectional_2018,tcheang_visual_2011}.

\begin{figure}[!hb]
	\centering
	\includegraphics[width=0.5\textwidth]{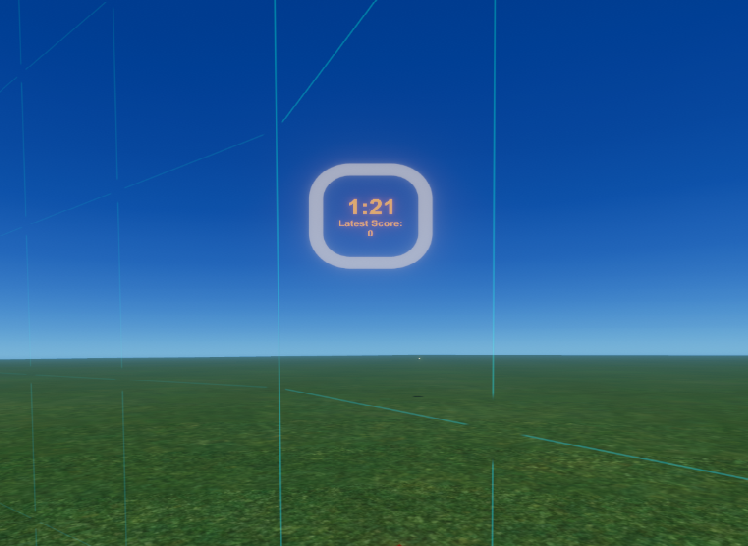}
	\caption{The visible blue grid marking the edge of a configured safe area.}
	\label{areaborder}
\end{figure}

In \ack, a visible grid barrier, shown in Figure~\ref{areaborder}, is displayed in the HMD when the participants approach the limits of a configured area in which they can safely move around. 
The grid barrier serves two purposes.
First and foremost, it prevents the participants from going out of the physical safe area, ensuring their safety.
It is advisable that a navigation task in \ack~be well confined to an area smaller than the safe area so that the participants will never encounter the barriers in the first place.
If they do view the barrier, it essentially functions as an extra landmark that informs about the boundary of an environment, which can induce significant bias in their navigational behavior \citep{cheung_estimating_2014,mou_defining_2013,bird_establishing_2010}.

Second, the barrier makes it possible to extend a navigable virtual space beyond the physical safe area, in case it is necessary for an experiment.
To do this, the participants hold a trigger button in the VR controller, which locks the VE in place. 
While the VE is locked, the participants' physical rotation in real space is not reflected in their virtual heading.
Thus, the participants appear to keep facing the same direction in the VE, despite physically turning to face away from the edge of the safe area. 
The grid barrier remains visible and in correct orientation in respect to the physical safe area, allowing the participants to reorient themselves before continuing. 
In order to minimize the motion sickness caused by the VE remaining still during the participants' physical rotation, \ack~blurs the VE during the rotation.
Unlike similar approaches used in the literature~\citep{williams2007exploring}, \ack~does not require the participants to rotate a fixed amount as long as they steer clear of the physical boundary.

Although this method of extending the virtual space can be practical, it must be used with caution because by physically rotating in the locked VE, the participants will most likely be forced to go through a mental process of dissociating real and virtual spaces once and realigning them after the physical rotation.
It is very probable that this process will have significant impact on the participants' mental and neural spatial representations, and in turn, on their subsequent navigational behavior.

\subsubsection{Teleportation}
\begin{figure}[!hb]
	\centering
	\includegraphics[width=0.5\textwidth]{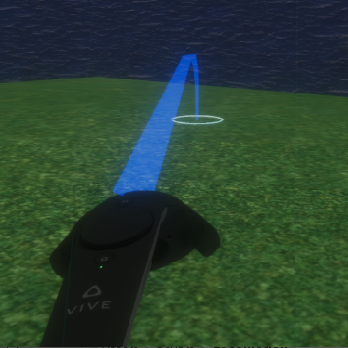}
	\caption{Teleportation locomotion with the teleportation marker visible.}
	\label{teleport}
\end{figure}

Teleportation locomotion differs from all of the other locomotion methods in that the participants never move through space, but instead teleport directly to their desired location some distance away. 
Before and after teleporting, the direction of the participants' body and head remain unchanged relative to the environment.
In \ack, the participants teleport by holding the trigger of a VR controller, which brings up the teleportation marker, as seen in Figure~\ref{teleport}. 
Then the participants place the marker on the desired teleportation target location by aiming the controller at the location.
Once the participants release the trigger, they are teleported to the marked location, given that the location is on the horizontal X-Z plane and clear of all collision regions around the objects in the environment.
A valid target location is indicated by the blue color of the marker, whereas invalid locations turn the marker red. 
Although teleportation is not a naturalistic method of locomotion, its use is increasingly common in VEs including those for spatial navigation research \citep{cherep_spatial_2020,kelly_teleporting_2020}.

\subsection{Goal Demarcation}
To accommodate a variety of experiments, \ack~offers two possible ways of demarcating the goal location for a navigation task: firstly, presenting static objects at goal locations (e.g., arrows, exclamation marks, or other similar objects) and secondly, showing dynamic objects like a buzzing fly that can give imprecise indication of the goal location. 
An example of the dynamic goal markers is available in the demonstration section below.

\subsection{Leader-board}
\begin{figure}[!htbp]
	\centering
	\begin{subfigure}{.5\textwidth}
	\includegraphics[width=0.95\linewidth]{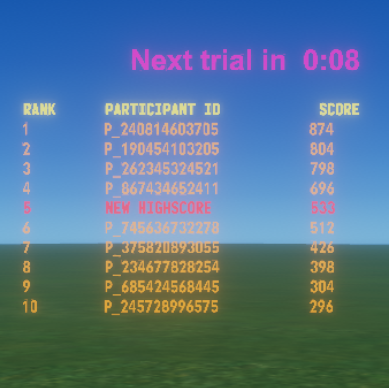}
	\caption{A new high-score is shown in red in the leader-board.}
\end{subfigure}%
\begin{subfigure}{.5\textwidth}
	\includegraphics[width=0.95\linewidth]{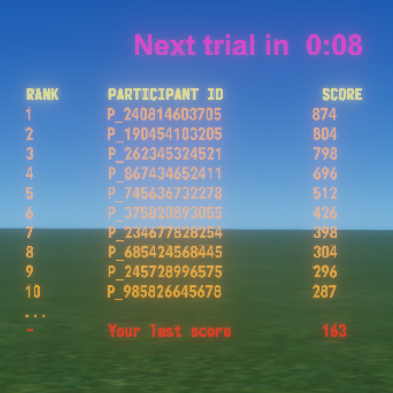}
	\caption{The latest score, which does not make it to the top ten, is shown at the bottom.}
\end{subfigure}
	\caption{Leader-boards inform participants about their performance in a given trial relative to their own and other participants' previous trials.}
	\label{fig:scoreboard}
\end{figure}

Gamification of learning has shown increased participant engagement be it for online programs~\citep{looyestyn2017does} or education~\citep{barata2013improving}.
Thus, as an optional feature, \ack~is equipped with a leader-board which provides a ranking based on scores obtained using Equation~\eqref{eqn:reward_func} or other equivalent equations implemented by experimentalists (Figure~\ref{fig:scoreboard}). 
\ack~keeps track of the scores and displays ten best scores in the leader-board. 
A new high-score is indicated with red font in the leader-board, while a score that was not high enough to get to the leader-board is shown at the bottom to illustrate the difference between the latest score and pre-existing scores. 
If participants are to carry out some practice or training trials first, it may not be appropriate to compare their scores against the pre-existing scores before they become fully familiarized with an experimental task.
In that case, it is possible to show a provisional ranking which is not integrated with the leader-board. 
For clarity, this is labeled with a red \textit{Practice} tag in the board.
Once the practice phase is finished, the actual scores of the participants are integrated into the leader-board that includes their own previous high scores.

While having a leader-board can motivate participants, it can also cause the conditions of an experiment to be different between the participants. 
As earlier participants obtain their place in the leader-board, they keep replacing lower scores on it. 
As such, it gets systematically more difficult for later participants to score high enough to make it to the top-ten scores of all time. 
Having a leader-board that is seemingly unreachable might provide a different motivation to the later participants than having an easily reachable one would. 
In order to ensure that each participant can have an equal experimental condition, \ack~offers two options.
First, because the leader-board is an optional feature, experimentalists can choose to remove it entirely.
Second, they can use a fake leader-board that behaves like a normal leader-board during the session of one participant, except that the changes to the board are not in fact stored to a log file. 
Once the next participant begins a session, the board reverts to its original condition, giving subsequent participants the same competitive challenge.
 
\subsection{Surveys}
\label{sec:survey}
\begin{figure}[!htbp]
	\centering
	\includegraphics[width=0.5\textwidth]{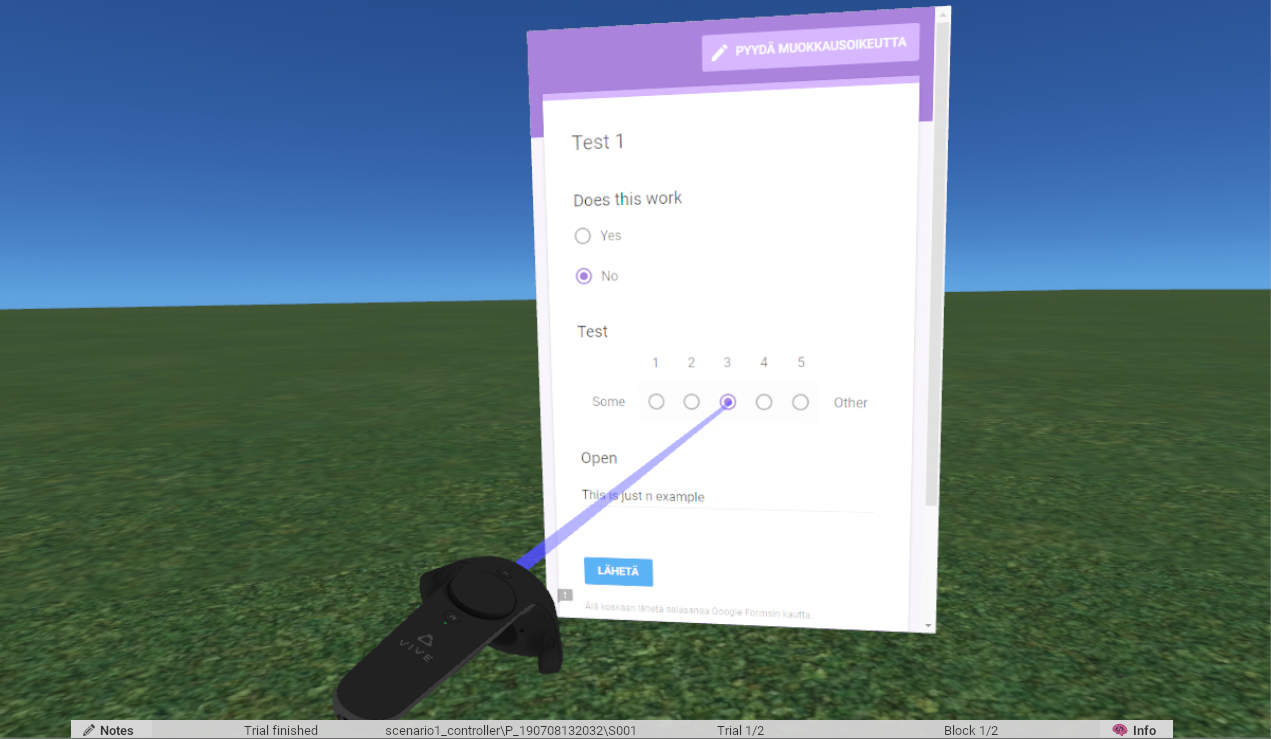}
	\caption{Filling in surveys whilst being immersed in VR using \ack.}
	\label{fig:survey}
\end{figure}

Often in behavioral studies, experimentalists would like participants to fill surveys for quality assurance or other purposes. 
Some of the most commonly used surveys for such studies using VR include the simulator sickness questionnaire~\citep[SSQ;][]{kennedy_simulator_1993} and the NASA task load index~\citep[NASA TLX;][]{hart2006nasa}. 
The SSQ studies the onset of simulator sickness symptoms like nausea or headache owing to being immersed in VR.
It contains 27 questions and the participants answer each of them using a scale ranging from none (0) to severe (3).
The NASA TLX is a survey for evaluating the workload of a given task utilizing six questions. 
Administering these and any other surveys has been made conveniently possible in \ack~(Figure~\ref{fig:survey}). 
The surveys are visible in an HMD to the participants and also on a desktop display to the experimentalists. 
While questions that have preset choices can be answered directly by the participants using the VR controller(s), questions that require free-form responses are to be typed in by the experimentalists on behalf of the participants. 

While \ack's survey system allows the experimentalists to administer surveys while keeping the participants immersed in VR, other systems typically require them to take off an HMD to answer the surveys.
Thus, if an experiment involves multiple sessions or blocks, each of which contains surveys, the participants need to be re-immersed in VR every time they remove the HMD and put it back on~\citep{schatz2017towards}.
This can be very cumbersome and make the participants feel uncomfortable, possibly inducing cybersickness.
An alternative could be that the experimentalists orally ask questions and fill in the surveys on behalf of the participants, but this can feel intrusive to the participants and reduce the sense of immersion in VR because the participants have to directly communicate with the experimentalists who do not belong to the virtual world \citep{bowman2002survey}.
\ack~remedies these issues by displaying the surveys in the HMD. 
To our knowledge, only \cite{grubel2017} and \cite{regal2018vrate} implemented a similar system previously.

\section{Demonstration}
To demonstrate various built-in functionalities of \ack~and its overall usability for human navigation experiments, we asked actual participants to perform a simple goal-directed navigation task in \ack.
In each trial, the participants navigated to a hidden goal that was indicated by a dynamic visual cue and received a feedback score on the leader-board.
The participants' navigation response was characterized by the total duration of the response, the residual distance to the goal at the end of the trial, the feedback score, and detailed time-course plots of their locomotion trajectory, exemplifying the types of data \ack~can record.
To implement the trial--block--session architecture, two blocks of trials were used, between which the environment and the dynamic cue were modified.
Two ways of deriving the feedback scores, which differentially weighted the response duration and the residual distance, were defined through \ack's settings files.
The participants also answered the SSQ and NASA TLX. This was to give a demonstration of \ack's intra-VR survey feature, and also to examine whether the participants tolerated the use of \ack~in the current experiment, which was of typical scale as a behavioral experiment.
\begin{figure}[!htbp]
  \centering
  \includegraphics[scale=0.55]{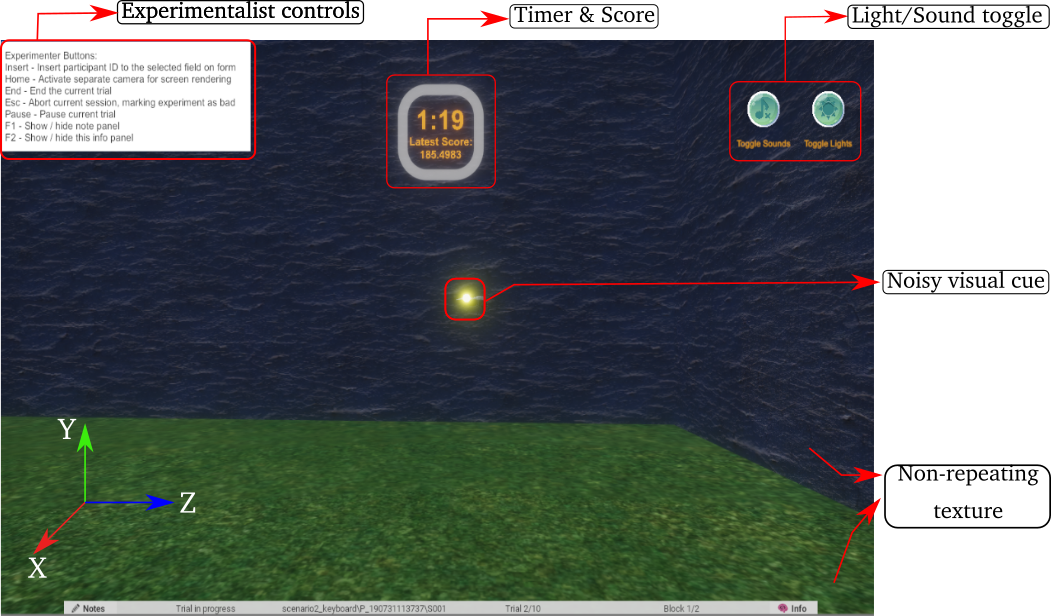}
  \caption{The default experimentalist view of an environment containing walls and floor with non-repeating textures. 
  A countdown timer and score window helps participants keep track of their performance. Light and sound can be toggled by experimentalists. The firefly serves as a noisy visual cue to guide the participants to a goal. 
  The light/audio toggle switches (top right) and experimentalist button descriptors (top left) are only visible to the experimentalists on a desktop running \ack, and are not made visible to the participants in an HMD. 
  Except for this difference, the participant and experimentalist views are the same.
  The arrows specifying X, Y, and Z axes in the lower-left corner are shown in this figure only; they appear neither in the participant nor experimentalist view.   
  See text for further details.} 
  \label{fig:env}
\end{figure}

\subsection{Method}
\subsubsection{Participants}
Twenty-four participants (15 males, 8 females, and 1 other) took part in this experiment. Twenty-three of them were students of Aalto University, and one was from the vocational university Metropolia, Finland. The mean age of the participants was $24.8\pm 2.9$ years.
The participants' educational background ranged from having graduated from high school to having a master's degree. 
All participants gave written informed consent to participate in the experiment and received a movie ticket in return for their participation. The protocol of the experiment was approved by the Aalto University Research Ethics Committee.

\subsubsection{Design and materials}
The participants were asked to navigate from start to goal positions within a virtual room of $10~\times10$~m with non-repeating textures as shown in Figure~\ref{fig:env}. 
The walls were $4$-m tall.
The participants navigated using the controller teleoperation method.
Specifically, they turned the head to face the direction they wanted to go and moved in that direction by pressing a button of a VR controller.
The participants' eye height in the virtual room was set at $1.36$~m, which approximately corresponded to their actual eye height while seated in a real room.
The participants first went through a block containing $15$ trials in which the room walls were visible.
Subsequently, they performed the same navigation task in the second block in which the walls were removed and the floor texture was extended to the horizon.
This block also consisted of $15$ trials.
The starting and goal positions were fixed across the trials as well as across the blocks.
Relative to the center of the room, in a left-handed coordinate system, the starting position was at ($4.5$~m, $4.5$~m) and the goal position was at ($-3$~m, $-1$~m).
At the starting position, the participants directly faced the hidden goal position with an orientation of $-135^\circ$ about the vertical Y-axis.
As a dynamic goal marker, a firefly buzzed around the goal position in such a way that its randomly fluctuating flying trajectory had its center directly over the goal position.
Specifically, in each frame, the fly's position along the horizontal X-Z plane was randomly sampled within the radii of 0.75 and 1.5 m from the goal in the first and second blocks, respectively.
The height of the fly along the Y-axis was randomly sampled in the range of $0.75$--$1.25~$m.
To make the fly move smoothly, its position was incremented with a step size of $5$~mm.
For a graphical presentation of X, Y, and Z axes, see Figure~\ref{fig:env}.

In this manner, the fly represented a noisy visual cue to guide the participants to the goal position.
That is, the exact goal position was never revealed to the participants, and instead they were told that the goal was somewhere inside the area delimited by the fly's trajectory.
Hence, the goal position was provided imprecisely to the participants via the noisy visual cue, and also the feedback score.

The participants were assigned to one of two groups, both of which received scores for their performance that depended on both navigation speed and accuracy, but with different weights. 
\textit{Time group} received feedback that put more importance on speed while the feedback in \textit{accuracy group} was weighted in favor of being as close as possible to the goal. 
The feedback provided to the participants was computed using Equation~\eqref{eqn:reward_func} with the constants shown in Table~\ref{table:score_groups}.
The score was presented to the participants at the end of each trial.

\begin{table}[!htbp]
\centering
\caption{Constants of the reward function for two participant groups.}
\begin{tabular}{l l l}
	\hline
	Constant & Time group value & Accuracy group value\\
	\hline
	$\alpha_1$ & $-0.05$ & 0.2 \\
	$\alpha_2$ & 0.2 & 1 \\
	$\beta_1$ & $-2$ & 0.5 \\
	$\beta_2$ & 6.2 & 3.4 \\
	\hline
\end{tabular}
\label{table:score_groups}
\end{table}

The participants were informed that they would be graded according to the time elapsed and residual distance to the actual goal. 
However, they were not told about the existence of the two groups or which group they belonged to. Instead, the participants of both groups were told that the scores obtained for the trials would be based on both speed and accuracy. 
In order to make the scores easier for the participants to follow, they were scaled up by a factor of 300 and their minimum value was set at 0 (i.e., no negative scores). 

\ack~was run on a high-performance Windows 10 personal computer with an Intel Core i5 processor, 32 gigabytes of random-access memory, and an Nvidia GTX 1070 graphics card. 
The HTC Vive Pro with a wireless extension was used for a VR HMD.
\ack~was used for presenting the VE and recording data for each frame (approximately 90 frames per second).
Specifically, the following log files were created for each participant per session: 

\paragraph{Environment settings} This file specified the size of the virtual room, the fly's trajectory (buzzing speed, minimum and maximum heights of flight, and a buzzing radius), and whether to remove or retain the bounding walls during the second block. Additionally, the links to survey forms were added here.
\paragraph{Locomotion settings} This log file specified the locomotion method used and its presets like traversal gain and rotation speed.
\paragraph{Scenario settings} This log file specified how many trials were to be presented in first and second blocks, the longest possible duration of a trial (with the maximum of 120 s), and start and goal locations.   
\paragraph{Participant particulars} The participant information as collected via the GUI (i.e., identification, age, gender, and highest qualification achieved) was recorded.
\paragraph{Movement logs} This log file recorded each participant's X and Z positions and rotation about the Y axis with time stamps. Owing to flexibility provided by \ack~to toggle lights and sounds even during a trial, the status of these parameters was also logged every frame in these logs. A new log was created per trial along with trial numbers.
\paragraph{Trial results} This log file assimilated component-wise and cumulative rewards per participant along with distance covered and time elapsed during a trial.
\paragraph{Notes} The experimentalists' notes during the experiment were recorded in this file. For instance, if some participants felt dizzy and opted for early termination, the particular session can be marked as bad and further details can be stored as notes for later use.

\subsubsection{Procedure}
The participants sat in the middle of a room that was clear of any obstacles.
At the outset of the experiment, the participants were asked to fill the SSQ to log their state of health before being immersed in VR.
Their age, gender, and the highest qualification achieved were also recorded using DeFINE's GUI (Figure~\ref{fig:GUI}).
An experimentalist then put an HMD on the participants' head (over the spectacles, as and when necessary) and handed them hand-held VR controllers.
The participants were run individually.

As soon as the participants had verbally confirmed to be ready, the first block was started by the experimentalist.
The participants began a trial by leaving the start position, using the controller teleoperation method to navigate, and ended their locomotion by pressing a key on a VR controller when they thought they had reached the goal position.
The goal was positioned diagonally across the other side of the room and remained unchanged across trials.
The participants then received a score from the trial in a leader-board (Figure~\ref{fig:scoreboard}).
The fake leader-board feature was used so that all participants performed the navigation task with the same competitive challenge.
The board was filled with made-up scores at the beginning, which were set low to give everyone a reasonable chance of making it to the top ten. Pilot testing was conducted to empirically determine how low should be low enough for this purpose.
The leader-board was displayed for 10 s (or until the participants pressed the ``End Trial'' key on the VR controller), and the room was automatically shown from the start position again for the next trial thereby resetting the scene to the exact same configuration for each trial.
The participants completed the trials at their own pace, until reaching the end of the block.
The participants were allowed to have a short break between trials. When necessary, the participants were able to skip a trial by pressing a controller key. 
At the end of the first block, the participants filled the NASA TLX in the \ack's form system (i.e., without taking off the HMD) using a 7-point Likert scale.
Upon having filled the form, the participants started the second block at their own input, prompted on the HMD. 
Once again the participants performed the trials at their own pace, until filling the NASA TLX one more time at the end of the second block. Filling the form completed the VR part of the experiment.

After taking off the HMD and the controllers, the participants filled the SSQ again to evaluate their simulation sickness after the exposure to the immersive VR.
In addition, the participants were invited to provide feedback about the experiment and \ack~by indicating the degree of agreement with each of the following five statements in a 5-point Likert scale: ``Instructions were easy to understand''; ``I understood what the score depended on''; ``moving in the VE was easy''; ``the walls in the practice phase were helpful''; and ``filling a form in the VE was easy''.

\subsection{Results}

Two participants from each group misunderstood task instructions and simply chased the fly rather than navigating towards the goal it indicated. 
This was determined through real-time observation of their locomotion patterns on the experimentalist's desktop (Figure~\ref{fig:env}).
Although it was not intended, this gave another demonstration of the utility of the experimentalist view that allowed the experimentalist to observe participants' responses as they took place during trials.
Due to this behavior, data from the four participants were excluded from analysis. 
The data presented in this section represent the results of the remaining $10$ participants per group, accounting for $20$ participants in total. 
In addition, in $0.7\%$ of all trials, the participants accidentally pressed the button to end the trial immediately after it had begun. 
These trials were also discarded for the analysis presented herewith. 

For each trial, the total elapsed time, the residual distance to the goal, and the score were derived as dependent measures from log files. 
The entire trajectory of a participant's locomotion was also reconstructed from recorded position data.
For each dependent measure, data points that were more than three standard deviations away from each participant's mean of each block were defined as outliers and removed from analysis. 
This resulted in removal of 1.17\% of trials on average.

\begin{table}[!b]
\centering
\caption{Means and standard deviations of the total elapsed time, the residual distance to the goal, and the score as a function of participant group and block.}
\begin{tabular}{l c c c c c}
	\hline
	 & \multicolumn{2}{c}{Time group} & & \multicolumn{2}{c}{Accuracy group}\\
	 \cline{2-3} \cline{5-6}
	 & First block & Second block & & First block & Second block\\
	 \hline
	Time (s) & 11.13 (5.93) & 7.81 (4.44) & & 10.97 (5.25) & 13.33 (10.77)\\
	Distance (m) & 0.43 (0.18) & 0.67 (0.32) & & 0.63 (0.34) & 0.80 (0.36)\\
	Score & 664.97 (239.02) & 718.87 (255.85) & & 624.77 (141.36) & 542.41 (160.26)\\
	\hline
	\multicolumn{6}{l}{\textit{Note.} Standard deviations are shown in parentheses.}\\ 
\end{tabular}
\label{table:descriptive_stats}
\end{table}

\subsubsection{Time, distance, and score}

Table~\ref{table:descriptive_stats} shows descriptive statistics of the dependent measures as a function of participant groups and blocks.
Overall, participants in the time group performed trials more quickly in the second block than in the first block, and those in the accuracy group showed the opposite pattern.
It is likely that this resulted from the types of feedback the participants received in each group---that is, the swiftness of a response was more heavily rewarded in the time group, whereas the closeness to the goal was given more emphasis in the accuracy group.
In terms of the residual distance to the goal, both groups performed worse in the second block, reflecting the fact that the navigation task was more challenging in the second block because of the lack of walls and decreased precision of the dynamic goal marker.
Capturing these outcomes, the scores increased in the time group and decreased in the accuracy group between the first and second blocks.

\begin{figure}[!ht]
	\centering
	\includegraphics[width=0.9\textwidth]{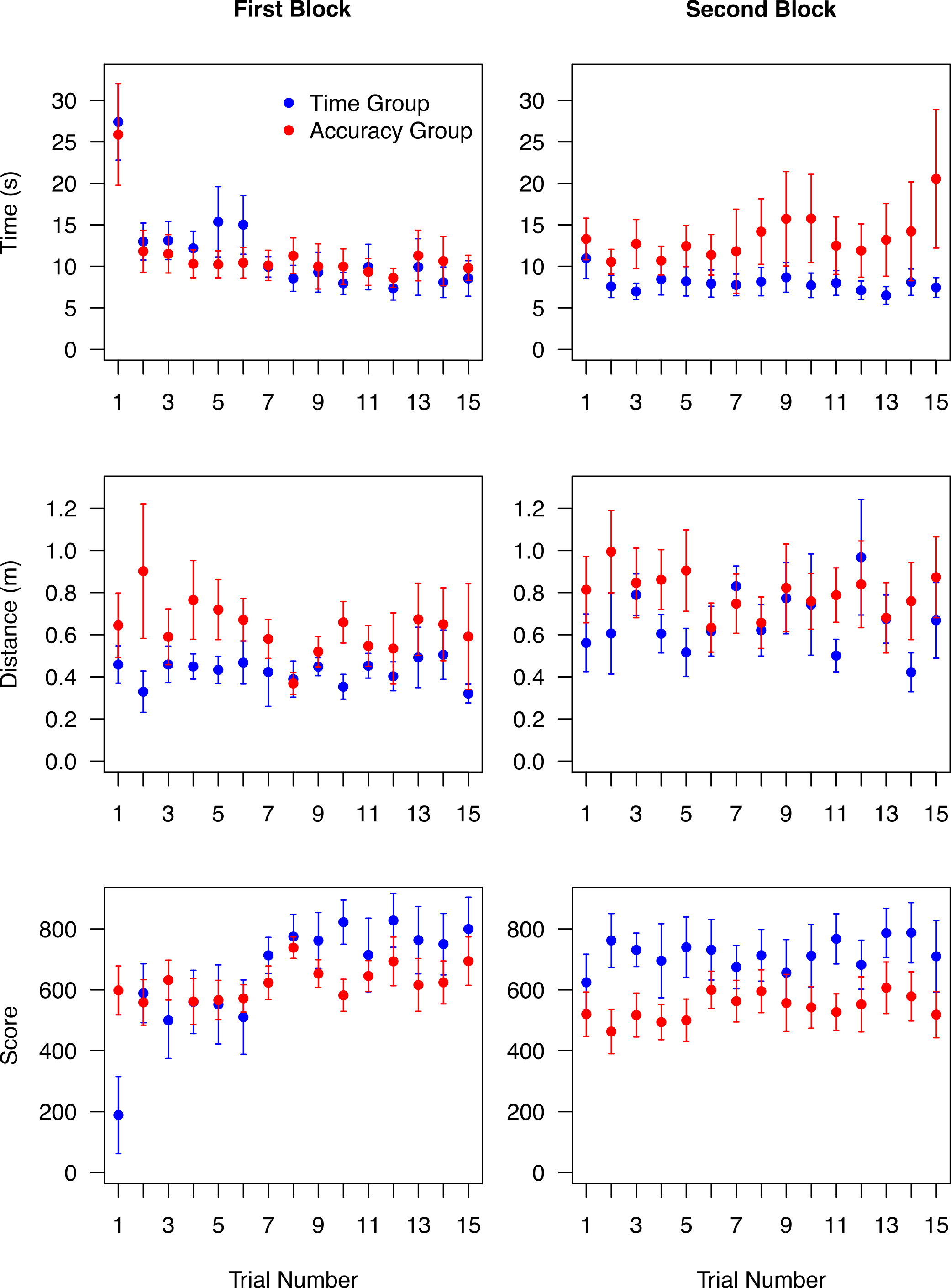}
	\caption{Mean times participants spent for performing each trial (top row), mean residual distances to the goal (middle row), and mean feedback scores the participants received (bottom row) as a function of participant group, block, and trial number. Error bars represent $\pm1$ standard error of the mean.}
	\label{time_and_dist}
\end{figure}

Figure~\ref{time_and_dist} shows the elapsed time, residual distance, and score in each trial, providing a more detailed picture of the participants' performance.
In terms of speed, the two groups performed similarly in the first block, but they differed in the second block.
Specifically, the time group maintained approximately the same speed throughout the block, performing the trials consistently quicker than the accuracy group.
This pattern suggests that the feedback scores affected the participants' navigation differently in the two groups.
On the other hand, with the reward parameters used in this experiment (Table~\ref{table:score_groups}), the effects of the scores were less clear on the accuracy of performance.
The participants in the accuracy group showed no visible improvement of accuracy in later trials, even though they received scores that rewarded accurate performance.
This might indicate that the parameters did not favor accuracy enough to elicit observable change in behavior within a block.
Between the blocks, it is also possible that the increase of task difficulty overrode the effects of the scores on navigation performance, as suggested by the overall lower scores in the second block.

To statistically examine the time, distance, and score data, they were analyzed separately by mixed analyses of variance (ANOVAs) in which block (first and second) was a within-participant factor and group (time and accuracy) was a between-participant factor.
Because the data plotted in Figure~\ref{time_and_dist} suggest that the first trial of the first block yielded rather different results (particularly in time and score), the ANOVAs were run with and without this trial.
The two sets of ANOVAs showed the same outcomes, and thus those including the first trial are reported here.
The main effect of block on distance was significant, $F(1, 18) = 10.52, p = 0.005, {\eta_G}^2 = 0.11$, reflecting the overall decrease of accuracy in the second block.
All the other main effects and interactions were not significant, $F\text{s}(1, 18) < 3.09, p\text{s} > 0.096, {\eta_G}^2\text{s} < 0.075$.

\subsubsection{Trajectory}

\begin{figure}[!hb]
	\centering
	\includegraphics[width=1\textwidth]{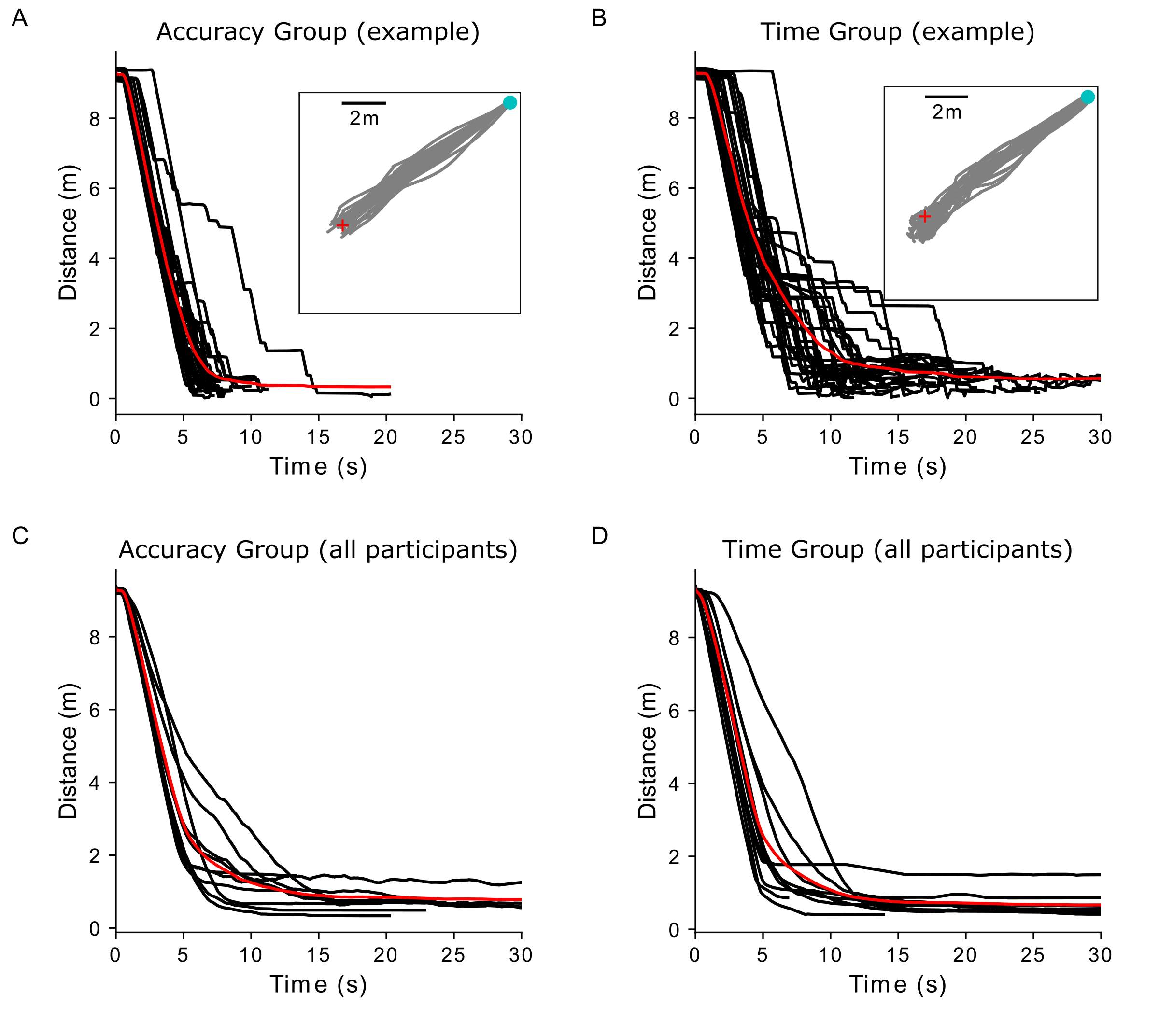}
	\caption{The residual distance to the goal as a function of participant group and time in a trial (first 30 s only). Panels A and B show sample participants' 30 trials (black lines), one participant from each group. The red lines represent the mean time courses of the residual distance derived from the 30 trials. To calculate the means, trials that lasted shorter than 30 s were extended by using the end values of each such trial. The insets display the corresponding trajectories of each participant from the start (cyan circle) to the goal (red cross). Panels C and D show the mean time courses of each participant (black lines) separately for the two groups. The red lines indicate group means. In calculating them, short time series were dealt with in the same way as in Panels A and B.}
	\label{trajectory}
\end{figure}

In addition to the time, distance, and score measures that aggregate participants' performance in a trial, \ack~can provide dynamic information about their movement within the trial.
To illustrate this feature, the time courses of the residual distance to the goal as well as sample participants' trajectories are plotted in Figure~\ref{trajectory}.
By visualizing the trajectories in this manner, researchers can gain additional insights into the way participants navigate in their experiments.
For example, by comparing panels A and B of Figure~\ref{trajectory}, it can be seen that the trajectories of the sample participant from the time group were more dispersed than those of the sample participant from the accuracy group.
Such an observation is easily possible in this visualization, but it is not readily available from the aggregate measures.

\subsubsection{Simulation sickness questionnaire (SSQ)}

Responses to the SSQ are summarized in Table~\ref{table:ssq}. As shown in the table, the participants scored very low not only before but also after exposure to the VE. Because the scores were very low overall, we used total raw scores for analysis, instead of deriving weighted scores for each sub-scale of the SSQ \citep{kennedy_simulator_1993}. The total raw SSQ scores were analyzed by a mixed ANOVA with exposure (before and after) as a within-participant factor and group (time and accuracy) as a between-participant factor. This ANOVA yielded no significant outcomes, $F\text{s}(1, 18) < 0.22, p\text{s} > 0.64, {{\eta_G}^2}\text{s} < 0.010$, suggesting that the SSQ scores did not differ between pre- and post-exposure to the VE as well as between the time and accuracy groups.
These results indicate that the use of \ack~did not induce any major symptoms of cybersickness.

Because non-significant results in an ANOVA do not necessarily constitute positive evidence for null hypotheses, we also conducted
a Bayes factor analysis to gauge the extent to which the data actually supported the claim that neither exposure nor group had an effect on the SSQ scores \citep{rouder_default_2012}.
When the null model was compared against the full model that included the main effects of exposure and group as well as the interaction between the two, it yielded a Bayes factor of 13.9.
This constitutes positive evidence for the null hypothesis \citep{kass_bayes_1995}, supporting the conclusion that doing the navigation task in \ack~(and which group each participant was in) did not cause cybersickness above and beyond what the participants had prior to navigating in the VE.

\bgroup
\def\arraystretch{0.9}
\begin{table}[!htbp]
\centering
\caption{Means and standard deviations of raw scores of the simulator sickness questionnaire (SSQ).}
\resizebox{\columnwidth}{!}{
\begin{tabular}{l c c c c c}
	\hline
	 & \multicolumn{2}{c}{Pre-exposure} & & \multicolumn{2}{c}{Post-exposure}\\
	 \cline{2-3} \cline{5-6}
	  & Time group & Accuracy group & & Time group & Accuracy group\\
	 \hline
	General discomfort & 0.1 (0.32) & 0.2 (0.63) & & 0.2 (0.42) & 0.3 (0.48) \\
	Fatigue & 0.4 (0.52) & 0.7 (0.95) & & 0.2 (0.42) & 0.6 (0.70) \\
	Boredom & 0.3 (0.67) & 0.1 (0.32) & & 0.3 (0.67) & 0.1 (0.32) \\
	Drowsiness & 0.5 (0.71) & 0.2 (0.42) & & 0 (0) & 0.2 (0.63) \\
	Headache & 0.1 (0.32) & 0 (0) & & 0.1 (0.32) & 0.2 (0.42) \\
	Eyestrain & 0.5 (0.53) & 0.5 (0.97) & & 0.7 (0.67) & 0.5 (0.97) \\
	Difficulty focusing & 0.2 (0.42) & 0.4 (0.70) & & 0.1 (0.32) & 0.3 (0.48) \\
	Salivation increase/decrease & 0.1 (0.32) & 0.2 (0.42) & & 0.1 (0.32) & 0 (0) \\
	Sweating & 0 (0) & 0.1 (0.32) & & 0.2 (0.42) & 0.2 (0.63) \\
	Nausea & 0 (0) & 0 (0) & & 0.2 (0.42) & 0.1 (0.32) \\
	Difficulty concentrating & 0.3 (0.48) & 0.3 (0.67) & & 0.1 (0.32) & 0.3 (0.48) \\
	Mental depression & 0.2 (0.42) & 0.2 (0.42) & & 0.2 (0.42) & 0.2 (0.42) \\
	Fullness of the head & 0.3 (0.48) & 0.4 (0.52) & & 0.3 (0.48) & 0.2 (0.42) \\
	Blurred vision & 0.1 (0.32) & 0.3 (0.48) & & 0.2 (0.42) & 0.4 (0.52) \\
	Dizziness with eyes open/closed & 0.1 (0.32) & 0.2 (0.63) & & 0.2 (0.42) & 0.3 (0.48) \\
	Vertigo & 0 (0) & 0 (0) & & 0.1 (0.32) & 0.1 (0.32) \\
	Visual flashbacks & 0 (0) & 0 (0) & & 0.3 (0.67) & 0.1 (0.33) \\
	Faintness & 0.1 (0.32) & 0 (0) & & 0 (0) & 0.1 (0.32) \\
	Breathing awareness & 0.4 (0.52) & 0 (0) & & 0.3 (0.48) & 0.1 (0.32) \\
	Stomach awareness & 0 (0) & 0.1 (0.32) & & 0.1 (0.32) & 0.1 (0.32) \\
	Loss of appetite & 0 (0) & 0 (0) & & 0 (0) & 0.1 (0.32) \\
	Increase of appetite & 0.1 (0.32) & 0.3 (0.48) & & 0.1 (0.32) & 0.4 (0.52) \\
	Desire to move bowels & 0.1 (0.32) & 0 (0) & & 0.1 (0.32) & 0 (0) \\
	Confusion & 0 (0) & 0.4 (0.70) & & 0.1 (0.32) & 0.1 (0.32) \\
	Burping & 0 (0) & 0 (0) & & 0 (0) & 0 (0) \\
	Vomiting & 0 (0) & 0 (0) & & 0 (0) & 0.1 (0.32) \\
	Others & 0 (0) & 0 (0) & & 0 (0) & 0 (0) \\
	\hline
	Total & 3.9 (3.51) & 4.6 (5.66) & & 4.2 (3.74) & 5.1 (4.12) \\
	\hline
	\multicolumn{6}{l}{\parbox[t]{\textwidth}{\textit{Note.} Standard deviations are shown in parentheses. The possible range of the total score was from 0 to 81.}}\\ 
\end{tabular}
} 
\label{table:ssq}
\end{table}
\egroup

\begin{table}[!htbp]
\centering
\caption{Means and standard deviations of scores of the NASA task load index (NASA TLX).}
\begin{tabular}{l c c c c c}
	\hline
	 & \multicolumn{2}{c}{Time group} & & \multicolumn{2}{c}{Accuracy group}\\
	 \cline{2-3} \cline{5-6}
	 & First block & Second block & & First block & Second block\\
	 \hline
	Mental demand & 3.0 (1.33) & 3.7 (1.64) & & 2.3 (1.25) & 3.0 (1.41)\\
	Physical demand & 1.7 (0.82) & 2.2 (1.14) & & 1.7 (1.25) & 2.7 (1.77)\\
	Temporal demand & 3.5 (1.35) & 4.1 (1.79) & & 2.9 (1.60) & 2.9 (1.85)\\
	Effort & 3.4 (1.58) & 3.6 (1.71) & & 2.8 (1.14) & 3.2 (1.32)\\
	Performance & 3.5 (1.43) & 2.6 (1.35) & & 3.9 (1.60) & 4.0 (1.70)\\
	Frustration level& 2.6 (1.71) & 3.2 (1.62) & & 1.9 (1.20) & 2.5 (1.43)\\
	\hline
	Total & 17.7 (6.38) & 19.4 (7.29) & & 15.5 (5.15) & 18.3 (7.89)\\
	\hline
	\multicolumn{6}{l}{\parbox[t]{0.8\textwidth}{\textit{Note.} Standard deviations are shown in parentheses. The possible range of the total score was from 6 to 42.}}\\ 
\end{tabular}
\label{table:nasatlx}
\end{table}

\subsubsection{NASA task load index (NASA TLX)}
Responses to each item of the NASA TLX ranged from one to seven, with smaller scores indicating lower task load.
As shown in Table~\ref{table:nasatlx}, participants generally indicated that doing the navigation task in \ack~required medium workload.
There was some variation of the scores between groups, blocks, and questions.
For example, the scores of the temporal demand question suggest that the time group felt stronger time pressure than the accuracy group, which is consistent with the feedback function that put emphasis on speedy response in the time group. 
In addition, scores in the second block tended to be higher than those in the first block, which corresponds to the fact that the task was made more difficult in the second block.
In line with these observations, a mixed ANOVA with block (first and second) and question (six questions of the NASA TLX) as within-participant factors and group (time and accuracy) as a between-participant factor yielded a significant interaction between question and group, $F(5, 90) = 2.96, p = 0.035, {\eta_G}^2 = 0.049, \epsilon = 0.66$ (this ANOVA was corrected for non-sphericity with the Greenhouse-Geisser method when appropriate).
The interaction between question and block as well as the main effect of question were also significant, $F(5, 90) = 3.31, p = 0.008, {\eta_G}^2 = 0.019$ and $F(5, 90) = 7.14, p < 0.001, {\eta_G}^2 = 0.11, \epsilon = 0.66$, respectively.
The main effect of block was not significant, $F(1, 18) = 3.58, p = 0.075, {\eta_G}^2 = 0.018$.
The interaction between block and group and the main effect of group were virtually non-existent, $F\text{s}(1, 18) < 0.36, p\text{s} > 0.55, {{\eta_G}^2}\text{s} < 0.010$, suggesting that overall, the two groups tolerated the workload of using \ack~in a similar way.

\subsubsection{Participant feedback on the experiment and \ack}
Scores of the participant feedback survey at the end of the experiment are summarized in Table~\ref{table:participant_feedback}.
Larger scores denote stronger agreement with the statements.
Overall, participants gave high scores, indicating that \ack~provided an easy-to-use interface for doing the navigation experiment.
The scores were analyzed by a mixed ANOVA in which statement (five statements in the survey) was a within-participant factor and group (time and accuracy) was a between-participant factor.
The main effect of statement was significant, $F(4, 72) = 6.51, p < 0.001, {\eta_G}^2 = 0.20$, which suggests that scores were reliably lower in the statement about the usefulness of walls than in the other statements.
The interaction between statement and group and the main effect of group were not significant, $F(4, 72) = 1.73, p = 0.15, {\eta_G}^2 = 0.062$ and $F(1, 18) = 0.18, p = 0.67, {\eta_G}^2 = 0.003$, respectively, suggesting that there was no overall difference between the groups in the way they responded to the feedback survey.

\begin{table}[!htbp]
\centering
\caption{Means and standard deviations of scores of the participant feedback survey.}
\begin{tabular}{l c c}
	\hline
	 &Time group & Accuracy group\\
	 \hline
	Clarity of instructions & 4.2 (0.63) & 4.1 (0.99)\\
	Score interpretation & 4.5 (0.71) & 4.0 (1.05)\\
	Ease of movement & 4.3 (0.67) & 4.2 (0.79)\\
	Usefulness of walls & 2.9 (1.37) & 3.3 (0.82)\\
	Ease of filling forms in \ack & 3.4 (0.97) & 4.2 (1.03)\\
	\hline
	Total & 19.3 (2.63) & 19.8 (2.62) \\
	\hline
	\multicolumn{3}{l}{\parbox[t]{0.8\textwidth}{\textit{Note.} Standard deviations are shown in parentheses. The possible range of the total score was from 5 to 25.}}\\ 
\end{tabular}
\label{table:participant_feedback}
\end{table}

\subsection{Discussion}

The purpose of this demonstration was to showcase the key functionalities of \ack---namely, its ability to set up an experiment in a trial--block--session structure, run goal-directed navigation trials in a stimulus--response--feedback loop, and collect both moment-by-moment and aggregate measures of participants' task performance.
In this experiment, the measures included the duration of a trial, the residual distance to the goal, the feedback score given to the participants, and the time course of the participants' positions within the trial.
It may be worth noting that among these measures, those that capture temporal information about the participants' locomotion (i.e., the trial duration and dynamic position data) can be of particular use in future studies because
past studies of goal-directed navigation in small-scale space tended to put emphasis on accuracy or precision of responses with little regard for how quickly they were carried out \citep[e.g.,][]{chen_cue_2017,chrastil_does_2014,harris_ageing_2012,yamamoto_homing_2014,yamamoto_medial_2014}.
However, it is important to consider the speed of the responses in evaluating their accuracy because there can be a trade-off relationship between them \citep{bogacz2010humans}. 
\ack~allows researchers to examine the speed and accuracy of navigation either in conjunction as in the current experiment or in isolation by setting the parameters of the reward function accordingly (e.g.,  $\beta_1 = 0$ makes the reward function exclusively focused on the accuracy). 

A defining feature of \ack~is that with its default settings, it provides feedback on participants' performance in each trial.
Results from this experiment suggested that by using differential weights on speed and accuracy of navigation in calculating feedback scores, \ack~has potential for eliciting different responses from the participants.
Although the behavioral measures of navigation were largely non-significant in the statistical analyses,
the effects of the feedback scores, particularly those on navigation speed, were implied in Figure~\ref{time_and_dist}.
The time group improved the speed of responses in early trials and kept the same speed during the rest of the experiment because this helped increase and then maintain the feedback scores.
On the other hand, the accuracy group appeared to care less for making a speedy response toward the end of the experiment, as the slower speed had little influence on the feedback scores in this group.
It is likely that the feedback scores were more effective in affecting the speed because of the specific way in which the current experiment was designed---that is, the participants were self-aware of the speed of their response, but the accuracy was never explicitly revealed to them, making it harder for the participants to improve the accuracy.
Importantly, this pattern is a result of one particular installation of \ack, and its architecture flexibly enables researchers to set up a suitable balance between the speed and accuracy according to the objectives of their studies.
For example, by giving heavier weights to accuracy in the feedback function, researchers can make feedback scores more directly informative about how well participants are reaching a goal.
Similarly, by demarcating the goal location more specifically by using different goal markers (e.g., a static marker or a dynamic marker with less variability) and environmental features (e.g., walls that provide spatial cues), researchers can run experiments in which focus is entirely on speed (i.e., accuracy is a given) or subtle changes in accuracy are scrutinized.

This experiment also examined the participants' experience in using \ack.
Results from the SSQ indicated that \ack~caused no major symptoms of cybersickness.
Considering that the participants repeatedly experienced sensory conflicts between their vision and body-based (i.e., vestibular and proprioceptive) senses due to the use of the teleoperation method, the absence of cybersickness is notable \citep{bos2008}.
The NASA TLX showed that the participants found doing the navigation task in \ack~moderately challenging but not unreasonably taxing.
In the feedback survey, the participants gave a positive evaluation to \ack~itself and the design of the experiment. 
Generally, these results did not differ between the two groups of the participants, suggesting that \ack~provided a versatile platform that accommodates different types of experiments.

In sum, this demonstration showed that by using \ack~in its default settings, we were able to run a human navigation experiment of typical scale, involving 24 participants and consisting of a single continuous session with two blocks of 15 trials each.
It exemplified an assortment of data types that \ack~can collect, which allow for detailed characterization of navigational behavior.
Both objective and subjective measures of the participants' experience indicated that they found \ack~easy to use and the navigation task it implemented well tolerable in terms of cybersickness and task workload, irrespective of the ways in which they carried out the navigation task (i.e., whether they were implicitly driven to perform it more quickly or accurately via feedback scores).
Together, these results validated \ack's capability and potential as a tool for investigating goal-directed navigation in humans under a variety of conditions. 
\section{Conclusions}
This paper presented the \framework (\ack) for studying goal-directed navigation in humans using VR. 
Although similar frameworks have already been developed \citep{Brookes2019,vasser2017vrex,commins_2019,machado2019new,wiener2019,geller2007,ayaz2008,ayaz2011,solway2013,grubel2017,bebko2020,starrett2020}, they are based on an open-loop stimulus--response architecture that omits performance feedback to participants.
\ack~distinguishes itself from the previous frameworks by implementing the closed-loop stimulus--response--feedback architecture as its core element (Figures~\ref{fig:loop} and \ref{fig:arch}).
The feedback is delayed by default in order to suit the needs of typical navigation experiments, but if necessary, it can be made real-time through relatively simple changes in \ack's code so that the stimulus--response--feedback loop is even more tightly closed.

%

As discussed in the introduction, the VR frameworks for navigation research mainly differ in whether they are geared toward ease of use by limiting their scope or wide applicability by providing general-purpose tools that demand technical skills of end-users.
In this spectrum, \ack~aims to position itself toward the ease-of-use end by focusing primarily on goal-directed navigation tasks and also by making it possible to set up an experiment mostly through GUIs and simple settings files (demonstrated in the video clips available online).
However, this does not mean that coding is absolutely unnecessary or impossible in \ack.
Indeed, some customization that goes beyond the GUIs and settings files is expected, and for this purpose the software is made open-source and its codebase is modularized.

The demonstration experiment showed the utility of \ack~as a platform for navigation research and its general friendliness to participants in VR experiments.
Additionally, this experiment demonstrated \ack's potential as a tool for testing hypotheses about the temporal aspects of navigational behavior.
The optional feature of seamlessly administering surveys within an HMD enhances the immersion of the participants in VR, thereby improving the quality of data collected via \ack.
The optional leader-board enables investigation of the effect of gamification on spatial navigation.
Previous studies have shown its impact in other domains of learning \citep{barata2013improving,looyestyn2017does}, but it is yet to be thoroughly explored for navigation-related applications \citep{coutrot_global_2018,coughlan_toward_2019}.
These out-of-the-box features of \ack, together with its customizability via the Unity software, open up many new possibilities for human navigation research. 
\newpage \section{Open Practices Statements}

The software used in the experiment reported in this article---the \framework (\ack)---is available at \url{https://github.com/ktiwari9/define-VR}.
The data and other materials for the experiment are available upon request. 
The experiment was not preregistered.

\bibliographystyle{spbasic}      
\bibliography{brm_define_2020}   

\end{document}